\documentclass[10pt,conference]{IEEEtran}

\usepackage{amsmath,amssymb,amsthm}
\usepackage{graphicx}
\usepackage{booktabs}
\usepackage{multirow}
\usepackage{enumitem}
\usepackage{placeins}
\usepackage{hyperref}
\usepackage[numbers,sort&compress]{natbib}
\usepackage[dvipsnames]{xcolor}

\hypersetup{colorlinks=true, linkcolor=MidnightBlue, citecolor=MidnightBlue, urlcolor=MidnightBlue}

\makeatletter

\def\@IEEEtablestring{table}
\long\def\@makecaption#1#2{%
\ifx\@captype\@IEEEtablestring%
\footnotesize\bgroup\par\centering
\@IEEEtabletopskipstrut{\normalfont\footnotesize #1.\nobreakspace #2}%
\par\addvspace{0.5\baselineskip}\egroup%
\@IEEEtablecaptionsepspace
\else
\@IEEEfigurecaptionsepspace
\setbox\@tempboxa\hbox{\normalfont\footnotesize {#1.}\nobreakspace\nobreakspace #2}%
\ifdim \wd\@tempboxa >\hsize%
\setbox\@tempboxa\hbox{\normalfont\footnotesize {#1.}\nobreakspace\nobreakspace}%
\parbox[t]{\hsize}{\normalfont\footnotesize \noindent\unhbox\@tempboxa#2}%
\else%
\hbox to\hsize{\normalfont\footnotesize\hfil\box\@tempboxa\hfil}%
\fi\fi}
\makeatother

\title{Quantum Parity Representations: Learnable Basis Discovery, Encoders, and Shadow Deployment}

\author{
\IEEEauthorblockN{Sang Hyub Kim, Oliver Knitter, Jonathan Mei, Claudio Girotto, Masako Yamada, Martin Roetteler, Chi Chen}
\IEEEauthorblockA{IonQ Inc., 4505 Campus Dr, College Park, MD 20740, USA\\ \{sang, oliver.knitter, jonathan.mei, claudio.girotto, yamada, martin.roetteler, chi.chen\}@ionq.co
}
}

\date{}

\begin{document}
\maketitle

\begin{abstract}
We study parity features as representations that can be evaluated entirely classically once the binary or quantized input representation and parity words are fixed, with particular interest in settings where labels depend on higher-order feature interactions or where discrete inference interfaces support perturbation robustness.
A parity feature is a signed product over selected bits of a binary input: once the participating bits are known, evaluation requires no quantum resources. Reaching a useful parity representation, however, requires solving two interrelated challenges. When the input is already parity-ready (a meaningful binary string), the main challenge is \emph{basis discovery}: selecting useful parity words from a combinatorial search space. When the input is not parity-ready, the main challenge is \emph{encoding}: constructing a binary vector on which parity computation is meaningful. 
We use hybrid quantum-classical training pipelines to address these challenges: learnable Pauli word selection for basis discovery, learned projection encodings for continuous embeddings, and sPQC-Parity for discrete inputs. On three native-binary parity tasks with 5--10 qubits, the learned parity basis improves mean accuracy by $23.9\%$ to $41.7\%$ over logistic-regression and support-vector baselines.
A model comparison shows that the improvement comes primarily from discovering the right parity basis, rather than from quantum moment computation at inference.
On five continuous text benchmarks, learned projection recovers much of the loss introduced by dimensionality reduction and fixed binarization, exceeding the full continuous baseline on CR, SST-2, and SST-5. On three encoding-limited discrete datasets, when compared with PCA-bin as the baseline, sPQC-Parity reaches $94.6\%$ improvement on mushroom, $3.0\%$ on splice, and matches PCA-bin on promoter. We also analyze inference robustness under binary or quantized inference, where rounding gives exact invariance below half the quantization step.
\end{abstract}

\section{Introduction}
\label{sec:intro}

Many machine learning tasks involve feature interactions that go beyond what linear or low-order models can directly represent.
When the label depends on a higher-order combination of features, a model must either enumerate candidate interactions explicitly or learn to detect them from data.
Parity features provide a natural representation for such interactions.
For a binary input string $\mathbf{b}\in\{0,1\}^n$, a parity feature is
\begin{equation}
f_S(\mathbf{b}) = (-1)^{\sum_{i\in S}b_i},
\label{eq:intro_parity}
\end{equation}
where $S\subseteq\{1,\ldots,n\}$ specifies which bits participate.
Once the participating bits are known, inference is a classical signed product---no quantum resources are required.
Parity features are also compatible with binary or quantized inference interfaces, which support a concrete form of inference robustness: after rounding back to the inference grid, any $\ell_\infty$ perturbation smaller than half the quantization step leaves the evaluated parity features unchanged.

Reaching a useful parity representation, however, requires overcoming two challenges that, depending on the input type, contribute differently.
If the input is already a meaningful bit string, the challenge is \emph{basis discovery}: selecting useful parity words from the $2^n$ possible Pauli-$Z$ strings.
Classical methods typically restrict the search to low-order candidates, which is effective when useful interactions are low order but misses the relevant word when the label depends on a higher-order interaction.
If the input is not already binary, the challenge is \emph{encoding}: constructing a binary vector on which parity computation is meaningful.
These two challenges correspond to two axes that determine practical usefulness: interaction order determines when parity features are the structurally appropriate representation, and encoder quality determines whether that representation survives compression strongly enough to matter in practice.

We use hybrid quantum-classical training pipelines to address these challenges.
Variational quantum circuits~\cite{peruzzo2014variational,havlicek2019supervised,biamonte2017quantum} naturally parameterize parity-type observables through Pauli-$Z$ expectations, and a differentiable relaxation of parity-word participation allows gradient-based search over the high-order word space without explicit enumeration.
Because the quantum stage is used only during training, inference runs entirely on classical hardware---the shadow deployment property~\cite{jerbi2024shadows}.

These two challenges inform the structure of the paper.
For parity-ready native-binary inputs, we introduce learnable Pauli word selection.
Instead of enumerating and ranking a fixed pool of parity words, we optimize continuous participation logits that relax the discrete choice of which qubits appear in each word.
For inputs that are not parity-ready, we study two encoding-side solutions.
sPQC-Parity learns a task-aligned binary vector for discrete or categorical datasets such as mushroom, splice, and promoter from PMLB~\cite{romano2021pmlb} using a simulated parameterized quantum circuit and learnable parity readout.
Learned projection encoding maps continuous embeddings to quantized binary vectors before parity computation.

\begin{figure*}[t]
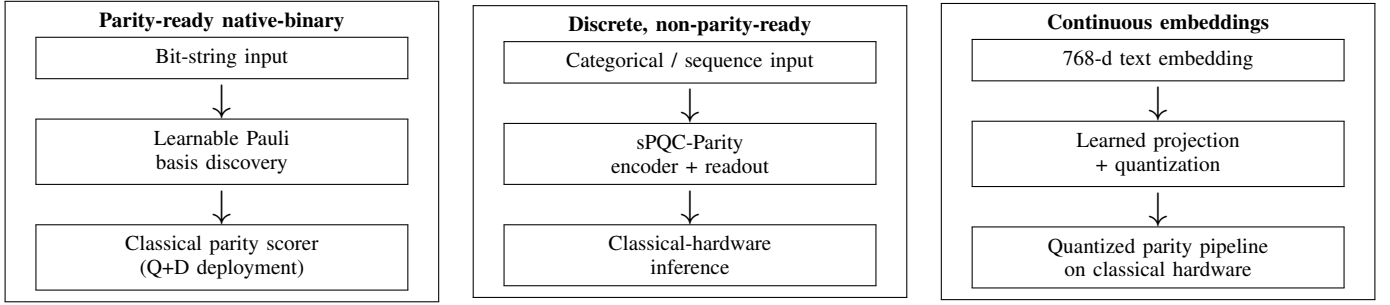

\centering
\footnotesize
\setlength{\fboxsep}{4pt}
\fbox{\begin{minipage}{0.30\textwidth}
\centering
\textbf{Parity-ready native-binary}\\[2pt]
\fbox{\parbox{0.85\linewidth}{\centering Bit-string input}}\\[2pt]
\centerline{\Large$\downarrow$}
\fbox{\parbox{0.85\linewidth}{\centering Learnable Pauli\\ basis discovery}}\\[2pt]
\centerline{\Large$\downarrow$}
\fbox{\parbox{0.85\linewidth}{\centering Classical parity scorer\\ (Q+D deployment)}}
\end{minipage}}
\hfill
\fbox{\begin{minipage}{0.30\textwidth}
\centering
\textbf{Discrete, non-parity-ready}\\[2pt]
\fbox{\parbox{0.85\linewidth}{\centering Categorical / sequence input}}\\[2pt]
\centerline{\Large$\downarrow$}
\fbox{\parbox{0.85\linewidth}{\centering sPQC-Parity\\ encoder + readout}}\\[2pt]
\centerline{\Large$\downarrow$}
\fbox{\parbox{0.85\linewidth}{\centering Classical-hardware\\ inference}}
\end{minipage}}
\hfill
\fbox{\begin{minipage}{0.30\textwidth}
\centering
\textbf{Continuous embeddings}\\[2pt]
\fbox{\parbox{0.85\linewidth}{\centering 768-d text embedding}}\\[2pt]
\centerline{\Large$\downarrow$}
\fbox{\parbox{0.85\linewidth}{\centering Learned projection\\ + quantization}}\\[2pt]
\centerline{\Large$\downarrow$}
\fbox{\parbox{0.85\linewidth}{\centering Quantized parity pipeline\\ on classical hardware}}
\end{minipage}}
\caption{\textbf{Training-time discovery and classical (shadow) deployment.} We consider three solution paths: basis discovery on parity-ready native-binary inputs, and encoding on discrete or continuous inputs that are not parity-ready. In each path, training produces a parity-compatible representation. The native-binary Q+D and learned projection paths instantiate classical inference without QPU calls, while sPQC-Parity probes the encoding side for discrete inputs.}
\label{fig:main_overview}
\end{figure*}

The experiments are organized to keep these questions separate.
Continuous text benchmarks such as CR from the SetFit collection~\cite{tunstall2022efficient}, SST-2 and SST-5~\cite{socher2013recursive}, AG News~\cite{zhang2015character}, and Emotion~\cite{saravia2018carer} measure how much information is lost when the full embedding is compressed and binarized, and how much a learned projection can recover.
Figure~\ref{fig:main_overview} summarizes the three paths studied in the paper. The paper makes four contributions.
\begin{enumerate}[leftmargin=1.5em]
\item We frame parity features as interaction-aware representations whose inference is entirely classical, and separate the problem into basis discovery and encoding, linked by two axes: interaction order and encoder quality.
\item We introduce learnable Pauli word selection, a differentiable relaxation of parity-word selection that searches over high-order parity bases using $O(Kn)$ trainable logits rather than explicit enumeration.
\item We develop two encoding-side parity pipelines: learned projection encoding for continuous embeddings, and sPQC-Parity for discrete inputs.
\item We use a model comparison to show that, on native-binary parity tasks, the main gain comes from quantum-guided basis discovery; classical data moments are enough for inference once the basis is known.
\end{enumerate}

\section{Parity representations}
\label{sec:parity_bottlenecks}

This section fixes the representation view used throughout the paper.
A parity word is a binary vector $\mathbf{s}\in\{0,1\}^n$.
Given a binary vector $\mathbf{b}$, it defines
\begin{equation}
p_{\mathbf{s}}(\mathbf{b})
=
\prod_{i=1}^{n}(-1)^{s_i b_i}
=
(-1)^{\mathbf{s}^{\top}\mathbf{b}}.
\label{eq:parity_word}
\end{equation}
The order of the word is its Hamming weight $\|\mathbf{s}\|_0$.
Low-order words describe individual bits or small interactions.
High-order words describe many-body interactions.

A deployed parity classifier has the form
\begin{equation}
\hat{y}
=
\mathrm{sign}\!\left(
\sum_{k=1}^{K} w_k (-1)^{\mathbf{s}_k^\top \mathbf{b}}
+
b_0
\right).
\label{eq:deployed_classifier}
\end{equation}
This model is classical once the words $\{\mathbf{s}_k\}_{k=1}^K$ and weights $\{w_k\}_{k=1}^K$ are fixed.
The quantum part of the pipeline is used before inference, either to discover the words or to construct the binary vector.

We use the term \emph{parity-ready} for inputs whose observed coordinates are already meaningful binary variables.
In that case, $\mathbf{b}$ is the input itself.
The hard part is deciding which parity words to use.
The search space contains $2^n$ possible words, and practical classical pipelines often restrict this space to low-order candidates.
That restriction is reasonable, but it can miss the exact word when the label depends on a higher-order interaction.

For inputs that are not parity-ready, the issue is different.
Before word selection can help, the model must map the original input $\mathbf{x}$ to a binary or quantized vector $\mathbf{b}$.
This is the encoding challenge.
It appears in discrete datasets whose raw categorical or sequence-valued representation is not already an aligned bit string, and in continuous embedding datasets where binarization can remove useful information.

The distinction matters for interpreting results.
On native-binary high-order parity tasks, improvements should be attributed to basis discovery.
On non-parity-ready tasks, improvements should be attributed primarily to better encoding.
The following sections keep those roles separate.

\section{Methodology}
\label{sec:method}

\subsection{Overview}
\label{sec:method_overview}

All results in this paper share the same inference target: a parity-compatible representation followed by a classical classifier.
They differ in how that representation is obtained.
For native-binary inputs, the binary vector is given, so the method focuses on discovering parity words.
For encoding-limited discrete datasets, sPQC-Parity learns a binary vector and parity readout jointly.
For continuous embeddings, learned projection encoding constructs a quantized vector before parity computation.

Throughout the paper, $\mathbf{b}$ denotes the binary vector on which parity features are evaluated.
In native-binary experiments, $\mathbf{b}$ is the observed input.
In quantum-moment experiments, $\mathbf{b}$ can be interpreted through measurement outcomes or Pauli expectations of a trained circuit.
In encoding experiments, $\mathbf{b}$ is produced by an encoder.
This notation keeps the parity readout fixed while allowing the source of the binary variables to change.

\subsection{Learnable Pauli word selection}
\label{sec:word_selection}

The basis-discovery problem is to select $K$ parity words
$\{\mathbf{s}_k\}_{k=1}^{K}$ with $\mathbf{s}_k\in\{0,1\}^n$.
A direct enumeration over all words is possible only at small $n$.
A more common practical strategy is to enumerate words up to a maximum Hamming weight $w_{\max}$ and rank them by a discriminative score.
This works when the useful interactions are low order, but it excludes all words above the cutoff.

We replace this discrete candidate selection with a differentiable relaxation.
Each word $k$ has a logit vector $\boldsymbol{\ell}_k\in\mathbb{R}^n$.
The soft participation of bit $i$ in word $k$ is $\sigma(\ell_{k,i})$, where $\sigma$ is the sigmoid.
For a binary vector $\mathbf{b}$, the relaxed parity feature is
\begin{equation}
f_k(\mathbf{b})
=
\prod_{i=1}^{n}
\cos\!\bigl(\pi\,\sigma(\ell_{k,i})\,b_i\bigr).
\label{eq:soft_parity}
\end{equation}
When $\sigma(\ell_{k,i})$ approaches either 0 or 1, this expression recovers the discrete parity word.
A non-participating bit contributes 1.
A participating bit contributes $(-1)^{b_i}$.
For intermediate values, the feature remains smooth, so gradients can update the participation logits.

The logits are trained together with the circuit parameters $\boldsymbol{\theta}$ using
\begin{equation}
\mathcal{L}
=
\mathcal{L}_{\mathrm{MMD}}(\boldsymbol{\theta})
-
\lambda
\mathcal{L}_{\mathrm{disc}}
\bigl(\boldsymbol{\theta},\{\boldsymbol{\ell}_k\}\bigr),
\label{eq:total_loss}
\end{equation}
where $\mathcal{L}_{\mathrm{MMD}}$ encourages the circuit distribution to match the data distribution~\cite{gretton2012kernel}.
The discriminative term is
\begin{equation}
\mathcal{L}_{\mathrm{disc}}
=
\frac{1}{K}
\sum_{k=1}^{K}
\mathrm{Var}_{c}
\left[
\mathbb{E}_{\mathbf{b}\sim p_c}
\bigl[f_k(\mathbf{b})\bigr]
\right],
\label{eq:disc_loss}
\end{equation}
where $c$ indexes classes.
This term rewards words whose expected parity values differ across classes.

At inference, each logit is thresholded:
\begin{equation}
s_{k,i}=\mathbb{1}\left[\sigma(\ell_{k,i})>0.5\right].
\end{equation}
The deployed model then uses hard parity words in Eq.~(\ref{eq:deployed_classifier}).
The number of word-selection parameters is $O(Kn)$.

\subsection{sPQC-Parity for non-parity-ready discrete inputs}
\label{sec:spqc}

The learnable word-selection method assumes that a useful binary vector already exists.
For many datasets, it does not.
sPQC-Parity addresses this case by learning an encoder and a parity readout together.
The method builds on the sPQC encoder introduced in~\cite{kim2024spqc} and adds learnable parity words at the readout.

Given an input $\mathbf{x}\in\mathbb{R}^d$, categorical features are one-hot encoded.
The vector is then zero-padded to dimension $2^{n_q}$, or reduced if needed, and normalized to prepare
\begin{equation}
|\psi(\mathbf{x})\rangle
=
\sum_b x_b |b\rangle / \|\mathbf{x}\|_2 .
\end{equation}
A hardware-efficient ansatz $U(\boldsymbol{\theta})$ is applied:
\begin{equation}
|\phi(\mathbf{x},\boldsymbol{\theta})\rangle
=
U(\boldsymbol{\theta})|\psi(\mathbf{x})\rangle.
\end{equation}
In the reported sPQC-Parity experiments, we use $n_q=14$ qubits, with dataset-specific circuit depth and feature count.
The feature for word $k$ is the expectation of the corresponding Pauli-$Z$ string:
\begin{equation}
f_k(\mathbf{x})
=
\langle \phi(\mathbf{x},\boldsymbol{\theta})|
Z^{\mathbf{s}_k}
|\phi(\mathbf{x},\boldsymbol{\theta})\rangle .
\label{eq:spqc_feature}
\end{equation}
A linear classifier maps the resulting $K$ features to class logits.

The training objective is
\begin{equation}
\mathcal{L}
=
\mathcal{L}_{\mathrm{CE}}
+
\alpha\Omega_{\mathrm{div}}
+
\beta\Omega_{\mathrm{sparse}}
+
\gamma\Omega_{\mathrm{disc}}.
\label{eq:spqc_loss}
\end{equation}
Here $\mathcal{L}_{\mathrm{CE}}$ is cross-entropy.
The diversity penalty discourages duplicate words,
\begin{equation}
\Omega_{\mathrm{div}}
=
\sum_{k\neq k'}
\left|
\langle
\sigma(\boldsymbol{\ell}_k),
\sigma(\boldsymbol{\ell}_{k'})
\rangle
\right|,
\end{equation}
and the sparsity penalty favors lower-weight words,
\begin{equation}
\Omega_{\mathrm{sparse}}
=
\sum_{k,i}\sigma(\ell_{k,i}).
\end{equation}
The discriminative term separates class means in the parity-feature space:
\begin{equation}
\Omega_{\mathrm{disc}}
=
-
\frac{1}{|\mathcal{P}|}
\sum_{(c,c')\in\mathcal{P}}
\left\|
\boldsymbol{\mu}_c-\boldsymbol{\mu}_{c'}
\right\|_2^2,
\end{equation}
where $\boldsymbol{\mu}_c=\mathbb{E}_{\mathbf{x}\sim c}[\mathbf{f}(\mathbf{x})]$.
The reported runs use $\alpha=1.0$, $\beta=0.01$, and $\gamma=2.0$.

Training uses a two-phase soft-to-hard schedule.
In the first phase, the temperature in $\sigma(\tau_t\ell_{k,i})$ is increased so that participation values move toward 0 or 1.
In the second phase, straight-through estimation uses hard selections in the forward pass and the soft surrogate in the backward pass~\cite{bengio2013estimating}.
This reduces the mismatch between training and the hard parity words used at inference.

\subsection{Learned projection encoding for continuous embeddings}
\label{sec:projection_encoding}

Continuous embeddings create a different encoding challenge.
The input is informative, but it is not a binary vector.
A naive sign or median-threshold binarization can discard signal before parity features are computed.
To measure this loss and test whether it can be recovered, we use a learned projection encoder.

Given a continuous embedding $\mathbf{x}$, the encoder applies a trainable linear map, followed by differentiable quantization during training and hard quantization at inference.
The output is a binary or low-bit vector.
Parity features are then computed on that vector and passed to a linear classifier.
This path is not meant to demonstrate quantum basis-discovery advantage.
Its role is to quantify the encoding challenge and test whether learned binarization is better than fixed PCA followed by sign thresholding.

\subsection{Model comparison}
\label{sec:swap}

The native-binary method combines two possible quantum ingredients.
The first is the \emph{basis}: which parity words are selected.
The second is the \emph{moment}: how feature values are computed for those words.
To separate them, we use a model comparison.

Let \textbf{D} denote the classical data-moment choice.
For the basis, D means top-$K$ words selected by empirical inter-class variance.
For the moments, D means empirical parity features computed directly from the input bit strings.
Let \textbf{Q} denote the quantum choice.
For the basis, Q means words obtained from the learned logits.
For the moments, Q means circuit expectations $\langle Z^{\mathbf{s}_k}\rangle$.
This gives four cells.

\begin{table}[t]
\centering
\caption{Model comparison: Moving \emph{down} a column isolates
the effect of the quantum-discovered basis; moving \emph{right} along a
row isolates the effect of quantum moment computation. The diagonal
$\text{D{+}D} \!\to\! \text{Q{+}Q}$ captures the total gain.}
\label{tab:swap_design}
\footnotesize
\setlength{\tabcolsep}{3pt}
\begin{tabular}{@{}lcc@{}}
\toprule
\textbf{} & \textbf{D moment} & \textbf{Q moment} \\
 & (emp.\ parity) & (circuit exp.) \\
\midrule
\textbf{D basis} (top-$K$ var.)  & D+D & D+Q \\
\textbf{Q basis} (learned logits) & Q+D & Q+Q \\
\bottomrule
\end{tabular}
\end{table}

The D+D cell is a fully classical parity baseline.
The Q+D cell uses the quantum stage only to discover the basis; inference uses empirical parities on raw bit strings.
The D+Q cell tests whether quantum moment computation helps when the basis is fixed classically.
The Q+Q cell is the full quantum basis and quantum moment configuration.
For inference, the most relevant cell is Q+D.
It keeps the quantum model in training and deploys only classical parity features.

\section{Results}
\label{sec:results}

We present results in the same order as the logic of the paper.
First, we test basis discovery on parity-ready native-binary tasks.
Second, we use model comparison to identify the source of the gain.
Third, we study encoders for discrete and continuous datasets.
Finally, we discuss robustness under binary or quantized inference.

\begin{table*}[t]
\centering
\footnotesize
\caption{\textbf{Main results summary.}
Each row reports a dataset-level quantitative outcome. Detailed numbers appear
in Tables~\ref{tab:native_parity}, \ref{tab:encoding_loss},
Fig.~\ref{fig:fgsm}, and Table~\ref{tab:swap_analysis}. The discrete
sPQC-Parity results are summarized here and discussed in
Section~\ref{sec:spqc_results}. For
native-binary tasks, the classical reference is the best overall
classical baseline; for discrete non-parity-ready and continuous
embedding tasks, both the full-feature and PCA-binarized (PCA-bin.)
references are shown, with $\Delta$ computed against the PCA-bin.\ baseline.}
\label{tab:main_summary}
\setlength{\tabcolsep}{2pt}
\begin{tabular}{p{0.13\textwidth}p{0.11\textwidth}p{0.10\textwidth}p{0.24\textwidth}p{0.13\textwidth}p{0.09\textwidth}}
\toprule
\textbf{Setting} & \textbf{Dataset} & \textbf{Method} & \textbf{Classical baseline} & \multicolumn{2}{c}{\textbf{Quantum}} \\
\cmidrule(lr){5-6}
 &  &  &  & \textbf{mean$\pm$std} & \textbf{$\Delta_{\mathrm{mean}}$} \\
\midrule
\multirow{3}{*}{\parbox{0.13\textwidth}{\centering Parity-ready\\native-binary}} & parity5 & \multirow{3}{*}{Learnable Pauli} & best 70.1 & \textbf{100.0$\pm$0.0} & \textbf{+29.9} \\
\cmidrule(lr){2-2}\cmidrule(lr){4-6}
 & synth.\ 3xor &  & best 69.1 & \textbf{93.0$\pm$4.6} & \textbf{+23.9} \\
\cmidrule(lr){2-2}\cmidrule(lr){4-6}
 & parity5\_5 &  & best 47.9 & \textbf{89.6$\pm$9.7} & \textbf{+41.7} \\
\midrule
\multirow{3}{*}{\parbox{0.13\textwidth}{\centering Discrete,\\non-parity-ready}} & mushroom & \multirow{3}{*}{sPQC-Parity} & \textbf{full 99.3}; PCA-bin.\ 84.8 & 94.6$\pm$3.2 & \textbf{+9.8} \\
\cmidrule(lr){2-2}\cmidrule(lr){4-6}
 & splice &  & \textbf{full 88.5}; PCA-bin.\ 72.8 & 75.8$\pm$2.9 & \textbf{+3.0} \\
\cmidrule(lr){2-2}\cmidrule(lr){4-6}
 & promoter &  & \textbf{full 78.1}; PCA-bin.\ 65.9 & 65.9$\pm$10.9 & +0.0 \\
\midrule
\multirow{5}{*}{\parbox{0.13\textwidth}{\centering Continuous\\embeddings}} & CR & \multirow{5}{*}{Learned proj.} 
& full 86.8; bin. 85.1 & \textbf{88.4$\pm$0.3} & \textbf{+3.3} \\
\cmidrule(lr){2-2}\cmidrule(lr){4-6}
& SST-2 &  & full 87.5; bin. 86.8 & \textbf{88.0$\pm$0.7} & \textbf{+1.2} \\
\cmidrule(lr){2-2}\cmidrule(lr){4-6}
& AG News &  & \textbf{full 83.7}; bin. 69.2 & 81.7$\pm$0.7 & \textbf{+12.5} \\
\cmidrule(lr){2-2}\cmidrule(lr){4-6}
& Emotion &  & \textbf{full 49.7}; bin. 41.4 & 46.0$\pm$2.2 & \textbf{+4.6} \\
\cmidrule(lr){2-2}\cmidrule(lr){4-6}
& SST-5 &  & full 40.0; bin. 40.4 & \textbf{43.2$\pm$0.6} & \textbf{+2.8} \\
\bottomrule
\end{tabular}
\end{table*}

\subsection{Basis discovery on native-binary parity tasks}
\label{sec:native_binary}

The native-binary tasks are the cleanest setting for basis discovery.
The inputs are already bit strings, and the labels depend on higher-order parity structure.
No upstream encoder is needed.
The remaining challenge is to discover which parity words should be measured.

Table~\ref{tab:native_parity} reports results on parity5, synthetic\_3xor, and parity5\_5.
All experiments use $K=128$ features, $K_{\mathrm{pool}}=256$, and 200 pretrain epochs.
Classical baselines include logistic regression and support-vector classifiers on continuous features, binarized features, and Bonferroni-selected parity features of orders 1--3.
The quantum column reports the Q+D deployment mode: the basis is discovered by the quantum training procedure, but feature values are computed as classical data moments at deployment.

\begin{table}[t]
\centering
\footnotesize
\caption{\textbf{Parity-ready native-binary tasks} (basis-discovery).
Learnable Pauli word selection with $K_{\mathrm{pool}}\!=\!256$.
Quantum results report mean$\pm$std and best seed over five seeds
$\{42,123,456,789,1024\}$.
$\Delta$ is quantum minus best classical baseline.
Quantum column reports the Q+D deployment mode (quantum-discovered basis, classical data moments).}
\label{tab:native_parity}
\setlength{\tabcolsep}{2.5pt}
\begin{tabular}{lrccccc}
\toprule
Dataset & $n_q$ & Best Cl. & Q mean$\pm$std & Q best & $\Delta_{\mathrm{mean}}$ & $\Delta_{\mathrm{best}}$ \\
\midrule
parity5       & 5  & 70.1 & \textbf{100.0$\pm$0.0} & \textbf{100.0} & $+29.9$  & $+29.9$ \\
synth.\ 3xor  & 10 & 69.1 & \textbf{93.0$\pm$4.6} & \textbf{100.0} & $+23.9$ & $+30.9$ \\
parity5\_5    & 10 & 47.9 & \textbf{89.6$\pm$9.7} & \textbf{100.0} & $+41.7$ & $+52.1$ \\
\bottomrule
\end{tabular}
\end{table}

On all three tasks, the quantum-discovered basis with classical data moments (Q+D deployment mode) substantially outperforms the best evaluated classical baseline, with mean-accuracy advantages of $23.9\%$ to $41.7\%$.
On parity5, all five seeds reach $100\%$ test accuracy, indicating that the learnable method recovers the planted 5-body parity reliably at the smallest native-binary scale considered here.
On synthetic\_3xor ($93.0\pm4.6$) and parity5\_5 ($89.6\pm9.7$), mean accuracy remains high and the best seed reaches $100\%$ in both cases, though seed-level variance indicates that the $n=10$ optimization landscape is harder to navigate reliably.
These results show that the learnable basis can recover higher-order parity structure that is missed by the evaluated classical baselines.

This should not be read as a claim that no classical method can recover the planted parity at these sizes.
At $n=10$, exhaustive enumeration is still possible.
The claim is narrower and more useful: when the classical candidate pool is restricted to practical low-order words, the learnable quantum-guided parameterization can move outside that pool and recover higher-order structure.

\subsection{Model comparison: the gain comes from basis discovery}
\label{sec:swap_results}

The model comparison tests whether the native-binary gain comes from discovering the right basis or from using quantum moments.
Table~\ref{tab:swap_analysis} shows the four cells on the two 10-qubit parity tasks.
We omit parity5 because all four cells reach $100\%$ once the full 5-qubit word space is small enough to search directly.

\begin{table}[t]
\centering
\footnotesize
\caption{\textbf{Model comparison for parity-structured tasks.}
Each cell uses a different combination of basis source (D = classical,
Q = quantum-learned) and moment source.
Mean over five seeds.
$\Delta_{\mathrm{best\,Q}}$ reports the gain of the best quantum cell
(Q+D or Q+Q, whichever is higher) over the fully classical baseline D+D.}
\label{tab:swap_analysis}
\setlength{\tabcolsep}{4pt}
\begin{tabular}{lccccc}
\toprule
Dataset & D+D & D+Q & Q+D & Q+Q & $\Delta_{\mathrm{best\,Q}}$ \\
\midrule
synth.\ 3xor & 59.4 & 59.4 & 93.0 & 93.0 & $+33.5$ \\
parity5\_5    & 46.5 & 46.5 & 89.6 & 89.6 & $+43.1$ \\
\bottomrule
\end{tabular}
\end{table}

On synthetic\_3xor, Q+D reaches $93.0\%$ while D+D reaches $59.4\%$.
On parity5\_5, Q+D reaches $89.6\%$ while D+D reaches $46.5\%$.
In both cases, Q+D and Q+Q are identical within rounding.
The quantum contribution is therefore not better moment computation.
It is basis discovery.

This distinction is important for inference.
Once the basis is found, the model does not need a QPU to evaluate features.
The Q+D configuration extracts the learned words and computes empirical parities directly on the input bit strings.
That is the deployment mode reported in the main native-binary results.

\subsection{Learned projection on continuous embeddings}
\label{sec:encoding_loss_results}

The clearest evidence for an encoding bottleneck appears on the continuous text benchmarks. These tasks use 768-dimensional sentence embeddings. The labels are not designed to require high-order parity structure, but the inputs must still be compressed and binarized before parity features can be computed.
The question is how much of that lost signal can be recovered by a learned low-dimensional vector.

\begin{table}[t]
\centering
\footnotesize
\caption{\textbf{Encoding loss on text benchmarks} (128 samples/class).
Full emb.\,= LR on the full 768-d SBERT embedding (no PCA).
PCA-bin.\,= PCA(14) followed by sign/median-threshold binarization and LR.
Learned proj.\,= pre-binarized $M$-bit learned projection within the
parity pipeline (clean accuracy, no attack; $\varepsilon=0$ from
Fig.~\ref{fig:fgsm}).
All results report mean$\pm$std over seeds $\{42,123,456\}$.}
\label{tab:encoding_loss}
\setlength{\tabcolsep}{3pt}
\begin{tabular}{lrccc}
\toprule
Dataset & $M$ & Full emb. & PCA-bin. & Learned proj. \\
\midrule
CR      & 1 & 86.8$\pm$2.4 & 85.1$\pm$1.0 & \textbf{88.4$\pm$0.3} \\
SST-2   & 1 & 87.5$\pm$1.3 & 86.8$\pm$0.6 & \textbf{88.0$\pm$0.7} \\
AG News & 1 & \textbf{83.7$\pm$0.3} & 69.2$\pm$2.7 & 81.7$\pm$0.7 \\
Emotion & 2 & \textbf{49.7$\pm$0.8} & 41.4$\pm$0.3 & 46.0$\pm$2.2 \\
SST-5   & 1 & 40.0$\pm$0.5 & 40.4$\pm$0.7 & \textbf{43.2$\pm$0.6} \\
\bottomrule
\end{tabular}
\end{table}

A fixed PCA-reduced binary vector can lose substantial accuracy.
Learned projection recovers much of that loss across all five benchmarks, and in some cases goes beyond the corresponding baselines.
On AG News, accuracy drops from $83.7\%$ with the full embedding to $69.2\%$ after PCA-bin, while learned projection reaches $81.7\pm0.7\%$.
On Emotion, learned projection improves over PCA-bin.\ by about five percentage points.
On CR and SST-2, it exceeds the full continuous reference ($88.4\%$ and $88.0\%$ vs.\ $86.8\%$ and $87.5\%$).
On SST-5, it reaches $43.2\%$, surpassing both the full embedding ($40.0\%$) and PCA-bin.\ ($40.4\%$).

These results are not intended as evidence of quantum basis-discovery advantage.
They show that encoding can be a real bottleneck, and that learned binarization can recover signal that fixed binarization loses.

\subsection{sPQC-Parity on encoding-limited discrete datasets}
\label{sec:spqc_results}

The discrete datasets show the same encoding bottleneck in a different form. Here the input is already discrete, but not yet parity-ready, so the model still needs to construct a useful binary vector before parity features can be applied.
The fair comparison is therefore against classical models using the same PCA-bin  interface.

Under that comparison, sPQC-Parity improves on mushroom
($94.6 \pm 3.2\%$ vs.\ PCA-bin.\ $84.8\%$, a $+9.8$~pp gain) and on
splice ($75.8 \pm 2.9\%$ vs.\ $72.8\%$, $+3.0$~pp).
On promoter, sPQC-Parity matches the PCA-bin.\ baseline
($65.9 \pm 10.9\%$ vs.\ $65.9\%$).
On all three datasets the full-feature LR baseline remains higher (mushroom $99.3\%$, splice $88.5\%$, promoter $78.1\%$), which indicates that the encoding bottleneck still limits both pipelines.

This pattern is consistent with the bottleneck view.
Where fixed PCA binarization discards useful structure, as in mushroom and splice, the learned quantum encoder recovers part of it.
Where the binarization loss is severe enough to collapse class separation, as in promoter, neither pipeline recovers the signal.

\subsection{Robustness under quantized inference}
\label{sec:robustness_results}

The primary inference-side robustness mechanism is interface-based and exact: if the deployed input is binary or quantized, rounding the perturbed input back to the grid removes any $\ell_\infty$ perturbation smaller than half the quantization step.
A secondary empirical effect is geometric: a low-dimensional parity readout can in practice produce smaller effective classifier weights, which may increase the perturbation size needed to cross the decision boundary.

\begin{figure*}[t]
\centering
\includegraphics[width=\textwidth]{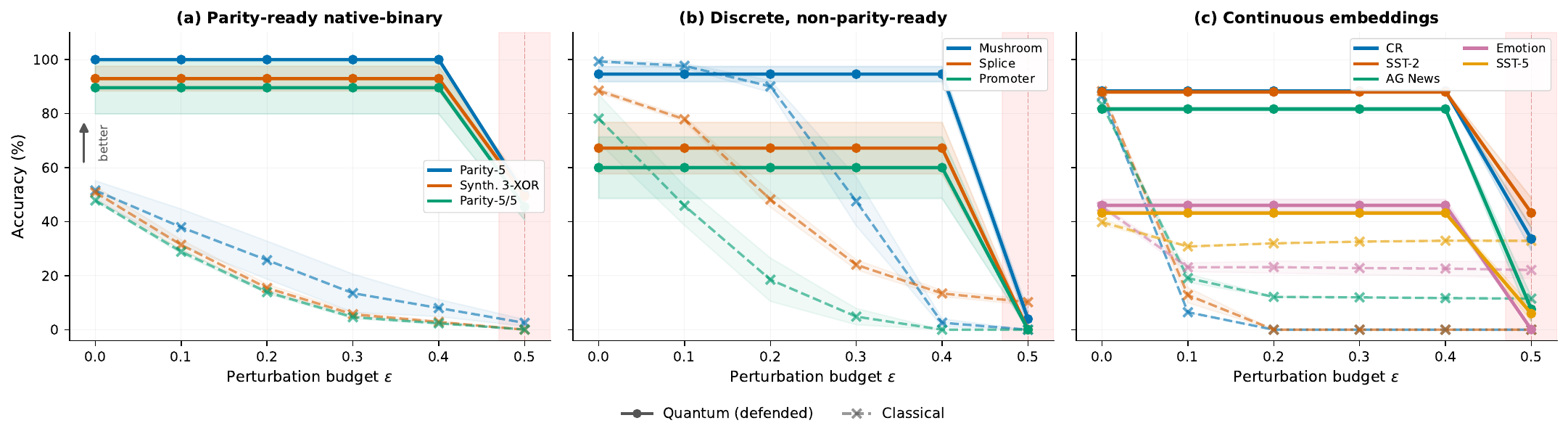}
\caption{\textbf{FGSM robustness on binary-input datasets.}
Accuracy (\%) under $\ell_\infty$ FGSM attack at perturbation budget
$\varepsilon$.
Solid lines = quantum parity model with input rounding to $\{0,1\}^d$;
dashed lines = classical LR baseline on the same feature space.
Shaded bands show $\pm$1 std over seed.
Rounding preserves \emph{exactly} the clean accuracy through
$\varepsilon < 0.5$ (all solid lines are flat);
at $\varepsilon = 0.5$ (red boundary) the mathematical guarantee
breaks and accuracy drops sharply.}
\label{fig:fgsm}
\end{figure*}

Fig.~\ref{fig:fgsm} summarizes FGSM accuracy for classical baselines, undefended parity models, and parity models with input rounding.

The undefended parity models can degrade sharply under FGSM.
The rounded models behave differently.
For binary inputs, rounding preserves the clean input for perturbations with $\varepsilon<0.5$.
As a result, the rounded parity pipeline keeps exactly the clean accuracy through $\varepsilon=0.3$ in the reported binary-input settings.
This should be interpreted as a property of the inference interface, not as a claim that the learned representation is intrinsically robust to all attacks.
The robustness evidence is strongest when inference is explicitly binary or quantized.

\section{Discussion}
\label{sec:discussion}

On native-binary high-order parity tasks, the challenge is basis discovery.
In this case the input is already a bit string, so the model does not need to learn an encoder, i.e., the model needs to find the right high-order parity words. This is accomplished with the learned Pauli-word method. And the model comparison confirms that this is indeed the source of the gain as
Q+D matches Q+Q, i.e., classical data moments are enough once the words are known.

On non-parity-ready inputs, the challenge is encoding. We presented results for continuous text benchmarks, where compression and fixed binarization can remove substantial signal and learned projection recovers much of it. In several cases exceeding the corresponding full continuous or PCA-bin. baselines.
The discrete results show the same challenge in a different form: sPQC-Parity recovers signal that PCA binarization loses on mushroom and splice, but reaches a ceiling on promoter where fixed binarization already collapses class separation.
Note that we do not claim a general quantum advantage over all classical algorithms. The learnable quantum-guided parameterization gives search access to higher-order words without explicitly enumerating the full parity basis.

Regarding inference, for native-binary Q+D and learned projection encoding, the model is shadow deployable, i.e., fully classical.
For sPQC-Parity, the reported discrete results are obtained by classically simulating a 14-qubit circuit. This is useful for studying whether the encoder can construct a parity-compatible vector, but it is not the same inference mode as extracting parity words and evaluating them directly on raw bit strings. Using sPQC-Parity as an inference method at scale would therefore require hardware execution, circuit compression, or extraction of a simpler classical representation.

We leave comparisons with stronger classical search procedures as future work. This includes randomized high-order parity pools, sparse Fourier methods, tree ensembles, kernel methods, and neural baselines.

\section{Experimental details}
\label{sec:methods}

\noindent\textbf{Datasets.}
We use three dataset families.
The native-binary parity tasks are parity5, synthetic\_3xor, and parity5\_5.
Their inputs are already in $\{0,1\}^n$, and their labels depend on high-order XOR structure.
The encoding-limited discrete datasets are mushroom, splice, and promoter from PMLB~\cite{romano2021pmlb}.
Their features are categorical or sequence-valued, so the model must learn or construct a useful binary vector.
The continuous embedding benchmarks are CR from the SetFit collection~\cite{tunstall2022efficient}, SST-2 and SST-5~\cite{socher2013recursive}, AG News~\cite{zhang2015character}, and Emotion~\cite{saravia2018carer}, using 768-dimensional frozen SentenceTransformer embeddings~\cite{reimers2019sentence} at 128 samples per class.
They measure loss from compressing and binarizing continuous representations.

\medskip
\noindent\textbf{Quantum configuration.}
Learnable selection experiments use $K=128$ features, $K_{\mathrm{pool}}=256$, and 200 pretrain epochs, with dataset-specific circuit depth and diversity weight summarized in Appendix~\ref{app:hyperparameters}.
The sPQC-Parity experiments use 14 qubits with dataset-specific circuit depth and feature count (mushroom: $L\!=\!6$, $K\!=\!128$; splice: $L\!=\!8$, $K\!=\!128$; promoter: $L\!=\!4$, $K\!=\!64$).
The selection schedule uses 100 epochs of temperature annealing followed by 100 epochs of hard forward selection with straight-through estimation.

\medskip
\noindent\textbf{Baselines.}
Classical baselines include logistic regression and support-vector classifiers on continuous features, binarized features, and Bonferroni-selected parity features of orders 1--3.
For text benchmarks, we compare full continuous embeddings, PCA-bin.\ binarization, and learned projection encoding.

\medskip
\noindent\textbf{Reporting.}
Native-binary experiments report mean, standard deviation, and best seed over $\{42,123,456,789,1024\}$.
Other result blocks state their seed sets in the table captions.
Mean accuracy is the primary comparison metric.
Best-seed accuracy is reported only as secondary context.

\section{Conclusion}
\label{sec:conclusion}

This paper studies parity representations for quantum-trained, classically evaluated learning.
The main idea is that parity features are useful when labels depend on higher-order interactions or when binary and quantized inference interfaces provide rounding-based inference robustness; in both cases, reaching a useful parity representation requires solving either a basis-discovery bottleneck or an encoding bottleneck.

On parity-ready native-binary tasks, learnable Pauli word selection discovers high-order parity bases that improve mean accuracy by $23.9\%$ to $41.7\%$ over the evaluated classical baselines.
The swap analysis shows that the gain comes from basis discovery rather than quantum moment computation.
Once the basis is known, classical data moments suffice for inference.

On inputs that are not parity-ready, the limiting factor is encoding.
The clearest encoding-side evidence comes from the continuous text benchmarks, where learned projection recovers much of the loss caused by dimensionality reduction and fixed binarization and, on several datasets, exceeds the corresponding full continuous or PCA-bin.\ baselines. On discrete non-parity-ready datasets, sPQC-Parity shows the same bottleneck in a different form, improving over the PCA-bin.\ on mushroom and splice and matching it on promoter.
Across these settings, the deployed model is best understood as a classical parity pipeline whose representation was discovered or constructed during training.

The result is a practical role for quantum models in classical machine learning: not as mandatory inference engines, but as tools for finding parity-compatible representations that can later be evaluated without a QPU.

\section*{Acknowledgment}
Generative AI tools (Anthropic Claude,  OpenAI Codex and Google Gemini) were used to assist with experiment scripting, data analysis support, and drafting or editing portions of the manuscript text. All scientific claims, experimental design, result verification, and interpretation were carried out and approved by the authors.

\appendices
\setlength{\textfloatsep}{6pt plus 1pt minus 2pt}
\setlength{\floatsep}{6pt plus 1pt minus 2pt}
\setlength{\intextsep}{6pt plus 1pt minus 2pt}

\section{Per-dataset hyperparameters}
\label{app:hyperparameters}

\begin{table}[ht]
\centering
\footnotesize
\caption{\textbf{Learnable selection (parity-ready native-binary) hyperparameters.}
All three datasets use $K\!=\!128$ features, $K_{\mathrm{pool}}\!=\!256$,
$200$ pretrain epochs, AdamW with cosine annealing.
Per-dataset values are selected on a held-out validation split drawn
from the training set at seed $42$ and fixed for all subsequent seeds.}
\label{tab:app_hparams_native}
\setlength{\tabcolsep}{4pt}
\begin{tabular}{lccccc}
\toprule
Dataset & $L$ & lr & dw & Pool init & Notes \\
\midrule
parity5      & 8 & 0.01 & 5 & top-$K$ classical & $n_q\!=\!5$ \\
synth.\ 3xor & 6 & 0.01 & 3 & random            & $n_q\!=\!10$ \\
parity5\_5   & 6 & 0.01 & 3 & random            & $n_q\!=\!10$ \\
\bottomrule
\end{tabular}
\end{table}

\begin{table}[ht]
\centering
\footnotesize
\caption{\textbf{Learned projection (text benchmarks, 14 qubits) hyperparameters.}
All runs use NCA-initialised projection, MI bit allocation,
RAA-MMD word discovery ($750+250$ epochs, best of 3 trials),
and Phase~2c end-to-end fine-tuning (25 epochs, $\tau$: $1.0\!\to\!0.1$).
Post-sel denotes whether post-selection refinement is enabled.}
\label{tab:app_hparams_learned_proj}
\setlength{\tabcolsep}{4pt}
\begin{tabular}{lcccc}
\toprule
Dataset & $M$ & lr & Post-sel & $L$ \\
\midrule
CR      & 1 & 0.005 & \checkmark & 6 \\
SST-2   & 1 & 0.005 &            & 8 \\
AG News & 1 & 0.020 & \checkmark & 6 \\
Emotion & 2 & 0.020 & \checkmark & 6 \\
SST-5   & 1 & 0.010 &            & 6 \\
\bottomrule
\end{tabular}
\end{table}

\noindent
All sPQC-Parity runs use $n_q\!=\!14$, $200$ total epochs ($100$ annealing
+ $100$ straight-through), learning rate $0.01$, Adam with cosine
annealing, $\alpha\!=\!1.0$, $\beta\!=\!0.01$, and $\gamma\!=\!2.0$.

\section{Per-seed results}
\label{app:per_seed}

On parity5, all five seeds reach $100\%$ because the relevant word space is small enough and the learned basis consistently recovers the planted parity.

\begin{table}[!htbp]
\centering
\footnotesize
\caption{\textbf{Per-seed accuracy (\%) on native-binary parity tasks.}
Q+D deployment mode, same configuration as
Table~\ref{tab:native_parity}.}
\label{tab:app_perseed_native}
\setlength{\tabcolsep}{3.5pt}
\begin{tabular}{lcccccccc}
\toprule
Dataset & S42 & S123 & S456 & S789 & S1024 & Mean & Std \\
\midrule
parity5      & 100.0 & 100.0 & 100.0 & 100.0 & 100.0 & 100.0 & 0.0 \\
synth.\ 3xor & 89.4 & 100.0 & 89.5 & 89.1 & 96.8 & 93.0 & 4.6 \\
parity5\_5   & 74.5 & 84.2 & 89.3 & 100.0 & 100.0 & 89.6 & 9.7 \\
\bottomrule
\end{tabular}
\end{table}

\begin{table}[!htbp]
\centering
\footnotesize
\caption{\textbf{Per-seed hard accuracy (\%) for learned projection on
text benchmarks} (14 qubits, best config per dataset, 3 seeds).}
\label{tab:app_perseed_learned_proj}
\setlength{\tabcolsep}{3pt}
\begin{tabular}{lccccc}
\toprule
Dataset & Seed 42 & Seed 123 & Seed 456 & Mean & Std \\
\midrule
CR      & 88.3 & 88.8 & 88.0 & 88.4 & 0.3 \\
SST-2   & 89.0 & 87.5 & 87.6 & 88.0 & 0.7 \\
AG News & 82.7 & 81.0 & 81.4 & 81.7 & 0.7 \\
Emotion & 44.3 & 49.2 & 44.6 & 46.0 & 2.2 \\
SST-5   & 43.8 & 43.4 & 42.4 & 43.2 & 0.6 \\
\bottomrule
\end{tabular}
\end{table}

\section{Learned Pauli word analysis}
\label{app:learned_words}

This appendix reports a post-hoc reproduction analysis of the learned words on parity5\_5, the largest native-binary benchmark in the paper.
The goal is to compare the learned words with the planted parity structure and with classical variance ranking.

The parity5\_5 label is defined by
\begin{equation}
y = \bigoplus_{i\in S^\star} b_i,
\end{equation}
with $S^\star=\{1,2,3,5,7\}$ using 0-indexed features.
The corresponding ground-truth word is
\[
\mathbf{s}^\star = \texttt{0111010100}.
\]

\begin{table}[!htbp]
\centering
\footnotesize
\caption{\textbf{Top-$5$ learned Pauli words on parity5\_5 vs.\
classical variance ranking (seed~42).} Both the learnable method and
full classical enumeration rank the planted word first; the D+D failure
comes from restricted search access rather than from the variance score
itself.}
\label{tab:app_learned_words}
\setlength{\tabcolsep}{3pt}
\begin{tabular}{l|ccc|ccc}
\toprule
Rank & Learned word & Ord. & $|w_k|$ & Classical word & Ord. & Score \\
\midrule
1 & \texttt{0111010100} & 5 & 4.06 & \texttt{0111010100} & 5 & 1.000 \\
2 & \texttt{1111100010} & 6 & 0.11 & \texttt{1010010101} & 5 & 0.035 \\
3 & \texttt{0000001011} & 3 & 0.10 & \texttt{1101011100} & 6 & 0.030 \\
4 & \texttt{1000001110} & 4 & 0.10 & \texttt{0101111101} & 7 & 0.030 \\
5 & \texttt{1111110010} & 7 & 0.09 & \texttt{1010000011} & 4 & 0.030 \\
\bottomrule
\end{tabular}
\end{table}

The learned method finds the ground-truth word at rank 1 in the reproduction analysis.
The classical variance score also identifies the same word when it is given access to the full enumeration of all non-trivial Pauli words.
This is an important point.
The classical scoring criterion is not the bottleneck at this scale.
The bottleneck is search access.
The restricted D+D pool used in the main comparison samples low-order words and excludes order-5 candidates, so it cannot contain the planted word.
The learnable method avoids that restriction by optimizing continuous participation logits and can move toward high-order words during training.

\FloatBarrier
\bibliographystyle{IEEEtran}
\bibliography{qce_2026_qtci}

\end{document}